\documentclass[aps,pre,singlecolumn]{revtex4}
\usepackage{amssymb}
\usepackage{graphicx}
\usepackage{colordvi}
\usepackage[dvipsnames]{xcolor}

\usepackage{amsmath}

\usepackage[utf8]{inputenc}

\newtheorem{theorem}{Theorem}[section]

\begin{document}

\title{Simulations of quantum dynamics with fermionic phase-space representations using numerical matrix factorizations as stochastic gauges} %

\author{François Rousse$^1$, Massimiliano Fasi$^{1,2}$, Andrii Dmytryshyn$^1$, M\aa rten Gulliksson$^1$, Magnus \"{O}gren$^{1,3}$}
\date{\today}

\affiliation{
$^1$School of Science and Technology, \"{O}rebro University, 70182 \"{O}rebro, Sweden \\
$^2$Department of Computer Science, Durham University, Stockton Road, DH1 3LE, UK \\
$^3$Hellenic Mediterranean University, P.O. Box 1939, GR-71004, Heraklion, Greece
}

\begin{abstract}
  The Gaussian phase-space representation can be used to implement quantum dynamics for fermionic particles numerically. 
	To improve numerical results, we explore the use of dynamical diffusion gauges in such implementations.
  This is achieved by benchmarking quantum dynamics of few-body systems against independent exact solutions.
  A diffusion gauge is implemented here as a so-called noise-matrix, which satisfies a matrix equation defined by the corresponding Fokker--Planck equation of the phase-space representation.
  For the physical systems with fermionic particles considered here, the numerical evaluation of the new diffusion gauges allows us to double the practical simulation time, compared with hitherto known analytic noise-matrices.
	This development may have far reaching consequences for future quantum dynamical simulations of many-body systems.

\end{abstract}

\maketitle

\section{Introduction} \label{sec:Introduction}

Quantum dynamics of few-body systems are benchmarked in order to study a new numerical method for diffusion gauges in the fermionic Gaussian Phase-Space Representation (GPSR)~\cite{Corney-fermionicII,Corney-fermionicI} with the goal to extend the practical usefulness of the method. 

A diffusion gauge is manifested by a time-dependent matrix $B \in \mathbb{C}^{n \times m}$ that satisfies the matrix equation 
\begin{equation} \label{eq:D_BBT}
D=BB^{T},
\end{equation}
where $D \in \mathbb{C}^{n \times n}$ 
represents the coefficients of second-order derivatives in the corresponding Fokker--Planck Equation (FPE).
By using the operator mappings of the positive-$P$~\cite{Drummond_JoPA_1980} for bosons and the fermionic Gaussian phase-space representation~\cite{Corney-fermionicII,Corney-fermionicI}, one can derive the FPE
\begin{equation}
\frac{\partial P\left(\vec{z}\right)}{\partial t}=-\sum_{j}\frac{\partial\left\{ a_{j}\left(\vec{z}\right)P\left(\vec{z}\right)\right\} }{\partial z_{j}}+\frac{1}{2}\sum_{j,k}\frac{\partial^{2}\left\{ D_{j,k}\left(\vec{z}\right)P\left(\vec{z}\right)\right\} }{\partial z_{j}\partial z_{k}}, \ \vec{z} \in \mathbb{C}^n,
\label{eq:FPE}
\end{equation}
for the time-dependent probability distribution $P\left(\vec{z}\right)$ of bosonic-, fermionic-, or mixed-bosonic-fermionic- systems containing terms with up to four creation- and annihilation-operators in a second quantized Hamiltonian.

We can rewrite the FPE~(\ref{eq:FPE}) as the Stochastic Differential Equation (SDE) in the It\^{o} interpretation~\cite{GardinerBook1}
\begin{equation} \label{eq:SDEs}
\dot{\vec{z}}=\vec{a}\left(\vec{z}\right) + \vec{b}(\vec{z}, \vec{\eta}), \ \ \ \vec{b}(\vec{z}, \vec{\eta})  = B\left(\vec{z}\right) \vec{\eta}, 
\end{equation}
where $\vec{\eta}$ is an $m \times 1$ Gaussian noise vector, and the dot denotes the derivative with respect to time, $d/dt$.
We need to construct a matrix $B \in \mathbb{C}^{n \times m}$ that satisfies the matrix equation~(\ref{eq:D_BBT})~\cite{GardinerBook1},
where $D$ is the matrix in~(\ref{eq:FPE}).
The freedom in $B$ defined by~(\ref{eq:D_BBT}) is known as a \textit{diffusion gauge}~\cite{Plimak_PRA_2001, Deuar_PRA_2002}.
This work focuses on the use of diffusion gauges that improves the durability in real time of the numerical implementation of GPSR.
Note that, in principle, the matrix $B$ can have any number of columns, and that a larger number of columns will increase the number of noise components in $\vec{\eta}$.

The stochastic averages over the complex phase-space variables $z_j$ in~(\ref{eq:FPE}) and~(\ref{eq:SDEs}) are directly related to first-order physical quantum-operators, $\hat{O}$, moments, i.e., $\langle z_{j} \rangle \leftrightarrow\bigl\langle \hat{O}_{j}\bigr\rangle $, and indirectly related to higher-order physical operator moments~\cite{Corney-fermionicI}.

In the absence of boundary corrections~\cite{Corney-fermionicI}, and impractically large sampling errors, the stochastic averages of the phase-space variables will approach the exact quantum mechanical expectation values as we take more and more trajectories, each trajectory being a particular realization of the SDE~(\ref{eq:SDEs}).
After a certain simulation time, however, boundary corrections may appear, 
or individual trajectories may tend to infinity, causing the numerical result to become unreliable from that point onward.
We refer to the time elapsed before the occurrence of these issues as the \textit{practical simulation time}.

When using imaginary time to represent the inverse temperature~$\sim T^{-1}$, similar boundary corrections hamper the calculations of fermionic groundstates, i.e., the $T \rightarrow 0$ limit.
This phenomenon was first documented for bosonic systems~\cite{Smith_PRA_1989, Gilchrist_PRA_1997}, and was then also observed  by researchers examining fermionic systems~\cite{Assaad_PRB_2005, Corboz_PRB_2008}, who managed to alleviate it by means of projection methods~\cite{Assaad_PRB_2005, Corboz_PRB_2008, Aimi_and_Imada_PRA_2007}.

However, for quantum dynamics~\cite{Ogren_EPL_2010, Ogren_CPC_2011, Corboz_bookchapter_2013}, i.e., real time evolution, boundary corrections have always been observed to be accompanied by clear signals, known as \textit{spiking trajectories}.

We propose and study a new numerical method for diffusion gauges in the fermionic Gaussian phase-space representation~\cite{Corney-fermionicII,Corney-fermionicI}.
We assess the new technique by benchmarking quantum dynamics of few-body systems against known exact solutions.

We also illustrate such spiking trajectories in calculations of fermionic quantum dynamics.
In particular, we show by numerical examples that the use of numerical diffusion gauges can delay the onset of spiking trajectories, and hence prolong the practical simulation time.
A numerical diffusion gauge is manifested by a matrix $B$, that for each discrete timestep in the implementation of the SDE, is a numerical solution to the matrix equation~(\ref{eq:D_BBT}).

Numerical diffusion gauges exhibit a second advantage: in practice, it may be cumbersome
to find an analytic solution $B$ to~(\ref{eq:D_BBT}) for a general Hamiltonian, although various specific examples are known~\cite{Ogren_EPL_2010, Ogren_CPC_2011, Corboz_bookchapter_2013}.

In the following sections, we discuss three examples of phase-space representations.
First, we introduce the well-known anharmonic oscillator for bosons~\cite{Quantum_Optics_Book}, for which, as mentioned, the role of diffusion gauges has been documented in the literature for at least 20 years.
Then, we describe a system governed by the Fermi--Hubbard Hamiltonian.
Finally, we discuss a mixed system that exemplifies the advantages of applying phase-space methods to a fermionic-bosonic Hamiltonian.
This is the so-called Fermi--Bose model~\cite{Ogren_JPA_2013}, which is here used to model the conversion from a molecular BEC to pairs of fermionic atoms.

These three examples of a bosonic-, fermionic-, and a mixed-system are of interest for example in quantum optics, solid state physics, and quantum atom-optics, respectively.
The method we consider, however, is more general and may find use for the exact quantum dynamics of many other Hamiltonians in different fields of applications.

\begin{figure}
\centering{}
\includegraphics[scale=0.42]{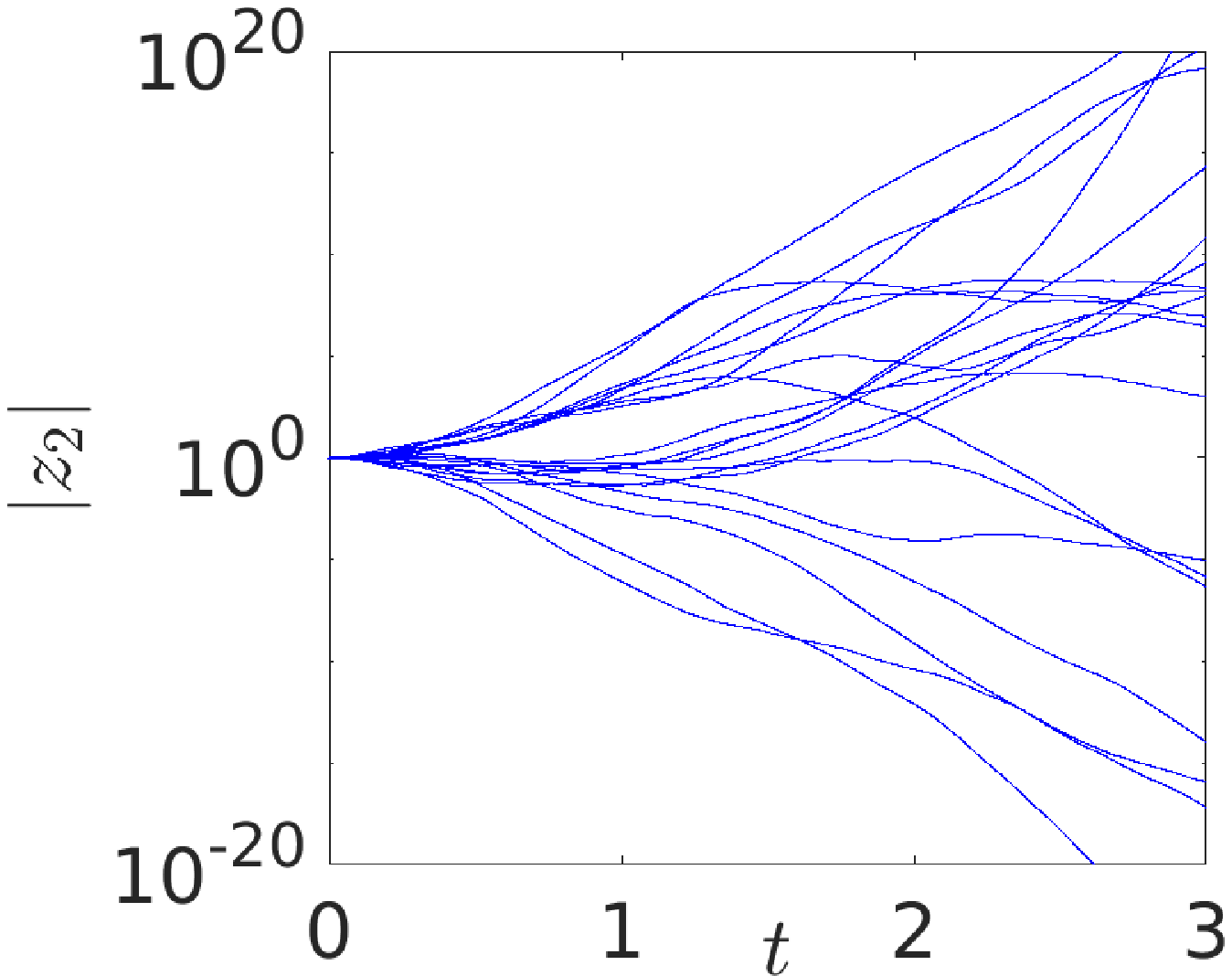}~\includegraphics[scale=0.42]{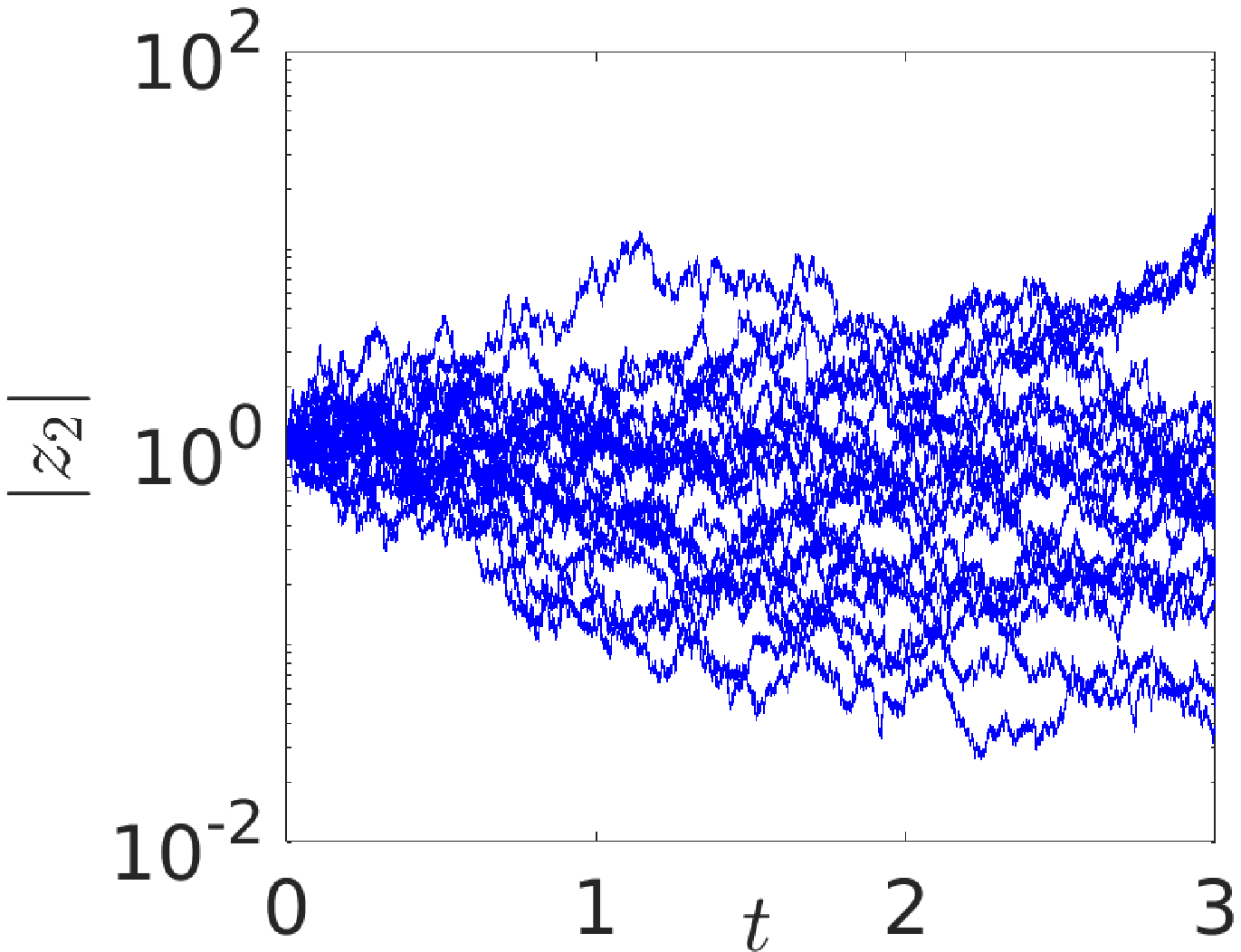}
\caption{(Colour online)
Positive-$P$ dynamics for $20$ individual trajectories with different diffusion gauges.
Left: Standard Positive-$P$, with the noise matrix $B$ of equation~(\ref{eq:ex_of_B_for_positive_P}).
Right:  ``optimal'' noise matrix, $B$ of equation~(\ref{eq:ex_2_of_B_for_positive_P}) with $\gamma \simeq 3.1985$, see~\cite{Deuar_PRA_2002}.
Note the different scales on the two logarithmic y-axes.
Parameters and initial conditions are $\omega=0,\: N=10^4,\: \chi=10^{-2}$ (i.e., $t_{\textnormal{opt}}=1/(\sqrt{N} \chi)=1$), respectively, according to equation~(\ref{positive_P_initial_conditions}), as in~\cite{Plimak_PRA_2001}.
}
\label{fig:ex_1_for_positive_P}
\end{figure}

\begin{figure}
\centering{}
\includegraphics[scale=0.6]{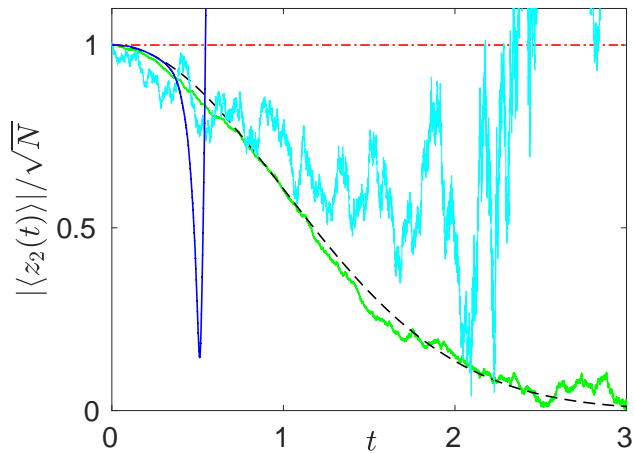}
\caption{(Colour online)
Benchmarking of the time-dependent correlation function
$ | \langle \hat{a}^\dagger(t)\hat{a}(0) \rangle | /N = | \langle z_2 (t) \rangle | / \sqrt{N} $
as a function of dimensionless
time~\cite{Plimak_PRA_2001}.
The dashed (black) curve describe the analytic result of
equation~(14) in~\cite{Plimak_PRA_2001}.
Solid (blue) curve is numerical results
from the phase-space method for $t \leq 0.5$ which is close to
the largest useful simulation time with the specific implementation
of~(\ref{eq:SDE_Bose-Hubbard}), while the solid (green) curve for $t \leq 3$ shows gauge-improved numerical results using instead a $B$ matrix from equation~(\ref{eq:ex_2_of_B_for_positive_P}).
We have sampled $10^{4}$, green solid, and $10^2$, cyan thin, to show the dependence on the number of trajectories.
The dashed-dotted (red) horizontal line describe the corresponding Gross-Pitaevskii
mean-field result, that do not vary with time in this case.
Parameters and initial conditions are the same as in figure~\ref{fig:ex_1_for_positive_P}.
}
\label{fig:observables_for_positive_P}
\end{figure}

\section{Examples of phase-space representations}

The first example, {\bf A}, which is meant as a simple introduction to diffusion gauges,  comes from early literature on the topic.
The following two examples, i.e., {\bf B} and {\bf C}, are representations for fermionic systems---we will use these to evaluate the performance of our numerical diffusion gauges in section~\ref{sec:numer-benchm-ferm}.

\subsection{A bosonic positive-$P$ phase-space representation, for an interacting bosonic quantum field}\label{section_positive-P}

As an introduction to the use of SDEs for exact quantum dynamics, we begin by revising the well-known time evolution of the one-mode Bose--Hubbard-like Hamiltonian (the Kerr oscillator~\cite{Quantum_Optics_Book})
\begin{equation}
\widehat{H}= \omega \hat{a}^{\dagger} \hat{a}   + \frac{\chi}{2} \hat{a}^{\dagger}\hat{a}^{\dagger}\hat{a}\hat{a}
.\label{eq:Bose-HubbardH}
\end{equation}
Note that here and in the remainder of this work we choose our units so that $\hbar=1$.

We formulate the equations for realizations with two $B$ matrices that are known analytically. 
If we combine equations~(\ref{eq:FirstOrderDerivatives_positive_P}) and~(\ref{eq:ex_of_B_for_positive_P}) in Appendix~\ref{app:Derivation_of_SDEs},
we obtain a specific system of SDEs $\dot{ \vec{z} } = \vec{a}(\vec{z}) +  \vec{b}(\vec{z}, \vec{\eta})$~\cite{Drummond_JoPA_1980}, which we can write elementwise as
\begin{equation}
\left[\begin{array}{c}
\dot{z}_1\\
\dot{z}_2
\end{array}\right]
= -i \left[\begin{array}{c}
\omega z_1 + \chi z_2 z_1^2\\
-\omega z_2 - \chi z_1 z_2^2\end{array}\right]  \\
+\sqrt{ i \chi }\left[\begin{array}{c}
 -i z_1 \eta_1 \\
 z_2 \eta_2 \\
\end{array}\right],
\label{eq:SDE_Bose-Hubbard}
\end{equation} 
where $ \vec{\eta} ^T= \left[ \eta_{1}, \eta_{2} \right]$ denotes uncorrelated real Gaussian noises with zero mean and unit variance.
We stress that in~(\ref{eq:SDE_Bose-Hubbard}) we follow the notation of~\cite{Gilchrist_PRA_1997} and~\cite{Steel_PRA_1998}, but the complex variables are defined differently by other authors~\cite{Plimak_PRA_2001}.
The specific noise terms in~(\ref{eq:SDE_Bose-Hubbard}) may follow from the one-parameter analytic diffusion gauge given in~(\ref{eq:ex_2_of_B_for_positive_P}) of Appendix~\ref{app:Derivation_of_SDEs}. 

If the noise terms in~(\ref{eq:SDE_Bose-Hubbard}) are set to zero ($\eta_{j}\equiv0$), then $z_2=z_1^*$ and we obtain the deterministic equation $i \dot{z}_1 = \left( \omega + \chi |z_1|^2 \right)z_1 $, which is equivalent to the time-dependent mean-field \textit{Gross--Pitaevskii} formalism \cite{BEC_book}.

For the initial condition
\begin{equation}
\vec{z}(0)^T=\sqrt{N}[1,1],
\label{positive_P_initial_conditions}
\end{equation}
where $N$ is the number of bosons, we calculate the dynamics of~(\ref{eq:SDE_Bose-Hubbard}) for different stochastic trajectories to obtain the average $\langle z_2(t) \rangle$.

In figure~\ref{fig:ex_1_for_positive_P}, we illustrate the different behavior of the dynamics for two choices of diffusion gauges: the standard positive-$P$ (left panel) and an improved (``optimal'') form~\cite{Plimak_PRA_2001} (right panel).
As is clearly seen from the 20 stochastic trajectories we report, which are realizations of the system of SDEs~(\ref{eq:SDE_Bose-Hubbard}), the variance of the standard positive-$P$ formulation (left figure~\ref{fig:ex_1_for_positive_P}), is several orders of magnitude larger than that of the ``optimal'' form~\cite{Plimak_PRA_2001} (right figure~\ref{fig:ex_1_for_positive_P}).
This different behavior will heavily influence the practical usefulness of averages obtained from such different trajectories.

We remark that the ``optimal'' form of the $B$-matrix used in figure~\ref{fig:ex_1_for_positive_P} was found by relying upon the existence of analytic solutions to certain correlations in the problem~\cite{Plimak_PRA_2001}.
Plimak, Olsen and Collett~\cite{Plimak_PRA_2001} demonstrated that changing the $B$-matrix so to dramatically reduce noise was enough to change the method from ``computable in principle'' to ``computable in practice''.
Although dependent on an analytic solution, this early example motivated an investigation of the role of the $B$-matrix in more difficult problems, for which analytic solutions are not known---a setting in which stochastic simulations can clearly be useful.

The stochastic average $\langle z_2(t) \rangle$ is here related to the quantum-operator average $\langle \hat{a}^\dagger(t) \hat{a}(0) \rangle$, given analytically in equation~(14) of~\cite{Plimak_PRA_2001}.
In figure~\ref{fig:observables_for_positive_P}, we use the analytic formula there to numerically illustrate the role of a diffusion gauge in this initial bosonic example.

\subsection{A fermionic Gaussian phase-space representation, for the Fermi--Hubbard model}
\label{sec:ferm-gauss-phase}

We study the time evolution
of a Fermi--Hubbard Hamiltonian with nearest-neighbour ($\left\langle i,j\right\rangle$) jumps and on-site interaction,
\begin{equation}
\widehat{H}=- J \sum\nolimits _{{\left\langle i,j\right\rangle },\sigma} \hat{c}_{{i},\sigma}^{\dagger}\hat{c}_{{j},\sigma} + U \sum\nolimits _{{j}}\hat{n}_{{j},\uparrow}\hat{n}_{{j},\downarrow}
,\label{eq:HubbardH}
\end{equation}
where $J$ is the hopping amplitude and $U$ the interaction strength.
We focus on a small two-site ($j=1$, $2$) chain in order to obtain detailed numerical comparisons for fermionic quantum dynamics.

First, we formulate the equations for realizations of the GPSR for the system in~\cite{Corboz_bookchapter_2013} with different choices of diffusion gauges.
We compare the results with independent numerical solutions in the number state representation (so-called exact diagonalization).

Combining equations~(\ref{eq:HubbAppFirstOrderDerivatives}) and~(\ref{eq:HubbAppBmatrix}) in Appendix~\ref{app:Derivation_of_SDEs}
gives the following set of SDEs, $\dot{ \vec{z} } = \vec{a}(\vec{z}) +  \vec{b}(\vec{z}, \vec{\eta})$~\cite{Corboz_bookchapter_2013}, or equivalently,
\begin{equation} \label{eq:SDE}
\left[\begin{array}{c}
\dot{z}_1\\
\dot{z}_2\\
\dot{z}_3\\
\dot{z}_4\\
\dot{z}_5\\
\dot{z}_6\\
\dot{z}_7\\
\dot{z}_8
\end{array}\right]
=i\left[\begin{array}{c}
J \left(z_2-z_3\right)\\
J \left(z_1-z_4\right)+U\left(z_5-z_8\right)z_2\\
J \left(z_4-z_1\right)+U\left(z_8-z_5\right)z_3\\
J \left(z_3-z_2\right)\\
J \left(z_6-z_7\right)\\
J \left(z_5-z_8\right)+U\left(z_1-z_4\right)z_6\\
J \left(z_8-z_5\right)+U\left(z_4-z_1\right)z_7\\
J \left(z_7-z_6\right)\end{array}\right] 
+\sqrt{ i U }\left[\begin{array}{c} 
i( \tilde{z}_1 z_1 \xi_{1}-z_2 z_3 \xi_{2} )
+
\tilde{z}_1 z_1 \xi_{3} - z_2 z_3 \xi_{4} \\
i( \tilde{z}_4 z_2 \xi_{2} - z_1 z_2 \xi_{1} )
+
\tilde{z}_1 z_2 \xi_{3} - z_2 z_4 \xi_{4} \\
i( \tilde{z}_1 z_3 \xi_{1} - z_3 z_4 \xi_{2} )
+
\tilde{z}_4 z_3 \xi_{4} - z_1 z_3 \xi_{3} \\
i( \tilde{z}_4 z_4 \xi_{2} - z_2 z_3 \xi_{1} )
+
\tilde{z}_4 z_4 \xi_{4} - z_2 z_3 \xi_{3} \\
i( \tilde{z}_5 z_5 \xi_{1}^* - z_6 z_7 \xi_{2}^* )
+
\tilde{z}_5 z_5 \xi_{3}^* - z_6 z_7 \xi_{4}^* \\
i( \tilde{z}_8 z_6 \xi_{2}^* - z_5 z_6 \xi_{1}^* )
+
\tilde{z}_5 z_6 \xi_{3}^* - z_6 z_8 \xi_{4}^* \\
i( \tilde{z}_5 z_7 \xi_{1}^* - z_7 z_8 \xi_{2}^* )
+
\tilde{z}_8 z_7 \xi_{4}^* - z_5 z_7 \xi_{3}^* \\
i( \tilde{z}_8 z_8 \xi_{2}^* - z_6 z_7 \xi_{1}^* )
+
\tilde{z}_8 z_8 \xi_{4}^* - z_6 z_7 \xi_{3}^* \\
\end{array}\right],
\end{equation}
where the notation $\tilde{z}_{j}\equiv1-z_{j}$ is used for the so-called hole-variables.
In~\eqref{eq:SDE}, we use the complex noise
\begin{equation}
  \xi_{j}=\dfrac{\eta_{1}^{\left(j\right)}+i\eta_{2}^{\left(j\right)}}{\sqrt{2}},
  \quad
  j=1,2,3,4,
\label{eq:ComplexNoises}
\end{equation}
where $\eta_{1}^{\left(j\right)}$ and $\eta_{2}^{\left(j\right)}$ denote real Gaussian noises with zero mean and unit variance.
The complex Gaussian noises $\xi_j$ obey the correlations
$$
\langle \xi_j (t) \xi_{j'} (t) \rangle = 0,
\qquad
\langle \xi_j (t) \xi_{j'}^*\ (t') \rangle = \delta_{j,j'} \delta_{t,t'}.
$$

If the noise-terms in~(\ref{eq:SDE}) are all set to zero ($\xi_{j}\equiv0$),
then we have a deterministic system that is equivalent to the time-dependent
Hartree--Fock formalism~\cite{RahavPRB2009} for the Hamiltonian~(\ref{eq:HubbardH}).

\subsection{A mixed Fermi--Bose phase-space representation, for dissociation of molecules}
\label{sec:mixed-fermi-bose}

As an example of a bosonic-fermionic system, we consider the dissociation of a molecular BEC of dimers into pairs of fermionic atoms, with the Hamiltonian
\begin{equation}
\widehat{H}=\hbar  \sum\nolimits _{ \mathbf{k} ,\sigma} \Delta_{\mathbf{k}} \hat{c}_{\mathbf{k},\sigma}^{\dagger}\hat{c}_{\mathbf{k},\sigma} - i \hbar \kappa \sum\nolimits _{\mathbf{k}} \left( \hat{a}^{\dagger} \hat{c}_{\mathbf{k},\uparrow}\hat{c}_{-\mathbf{k},\downarrow}  -   \hat{c}_{-\mathbf{k},\downarrow}^{\dagger}  \hat{c}_{\mathbf{k},\uparrow}^{\dagger} \hat{a} \right)
,\label{eq:MBEC_H}
\end{equation}
where $\Delta_{\mathbf{k}}=\Delta + \hbar {\mathbf{k}}^2 /(2m)$ is a parameter for the kinetic energy of the free atoms, translated with respect to the energy for the bound molecular state~\cite{Ogren_EPL_2010}, and $\kappa$ is the strength of the atom-molecular coupling.

With only one momentum mode $\mathbf{k}=1$, and symmetric initial conditions with respect to the spin-variable $\sigma$, we only need one phase-space variable for the fermionic normal moment $ \langle \hat{c}_{{\mathbf{k}},\sigma}^{\dagger}\hat{c}_{{\mathbf{k}},\sigma}  \rangle    \leftrightarrow  \langle   z_1  \rangle $ and one for the anomalous moment $  \langle \hat{c}_{{\mathbf{k}},\uparrow} \hat{c}_{{-\mathbf{k}},\downarrow}  \rangle  \leftrightarrow   \langle z_2 \rangle $, while the bosons are again mapped via the positive-$P$ representation (see section~\ref{section_positive-P}).

Here, we write down the minimal system with the 5 phase-space variables ($z_1=n_1$, $z_2=m_1$, $z_3=m^+_1$, $z_4=\alpha$, $z_5=\alpha^+$, compare to~\cite{Ogren_EPL_2010}) in explicit form.
From equations~(\ref{eq:FirstOrderDerivatives_molecular_dissociation}) and~(\ref{eq:ex_of_B_for_FBM}) in Appendix~\ref{app:Derivation_of_SDEs}, we have the system of SDEs $\dot{ \vec{z} } = \vec{a}(\vec{z}) +  \vec{b}(\vec{z}, \vec{\eta})$~\cite{Ogren_EPL_2010}, where
\begin{equation} \label{eq:molecular_dissociation}
\left[\begin{array}{c}
\dot{z}_{1}\\
\dot{z}_{2}\\
\dot{z}_{3}\\
\dot{z}_{4}\\
\dot{z}_{5}
\end{array}\right] =
\left[\begin{array}{c}
\kappa \left( z_3 z_4 + z_2 z_5 \right)\\
  -2i\Delta_1 z_2 + \kappa z_4 \left( 1 - 2 z_1 \right) \\
   2i\Delta_1 z_3 + \kappa z_5 \left( 1 - 2 z_1 \right) \\
-\kappa z_2\\
-\kappa z_3\\
\end{array}\right]
+ \sqrt{\kappa}
\left[\begin{array}{c}
z_1 \left( z_2 \xi_1^* + z_3 \xi_2^* \right) \\
z_2^2 \xi_1^* - z_1^2 \xi_2^*\\
-z_1^2 \xi_1^* + z_3^2 \xi_2^*\\
 \xi_1\\
 \xi_2
\end{array}\right].
\end{equation}
If the noise-terms in~(\ref{eq:molecular_dissociation}) are all set to zero ($\xi_{j}\equiv0$), then we have a deterministic system that is equivalent to the time-dependent pairing-meanfield formalism~\cite{PMFT_Jack_and_Pu_PRA_2005} for the Hamiltonian~(\ref{eq:MBEC_H}).

\section{Numerical benchmarking of fermionic systems} 
\label{sec:numer-benchm-ferm}

Now we describe the numerical experiments used to benchmark the fermionic systems presented in sections~\ref{sec:ferm-gauss-phase} and~\ref{sec:mixed-fermi-bose} above, we discuss the results we obtained, and we show how the numerical diffusion gauge based GPSR method we propose improves on existing alternatives found in the literature.

\subsection{Results for the Fermi--Hubbard model}
%
\begin{figure}
\centering{}
\includegraphics[scale=0.62]{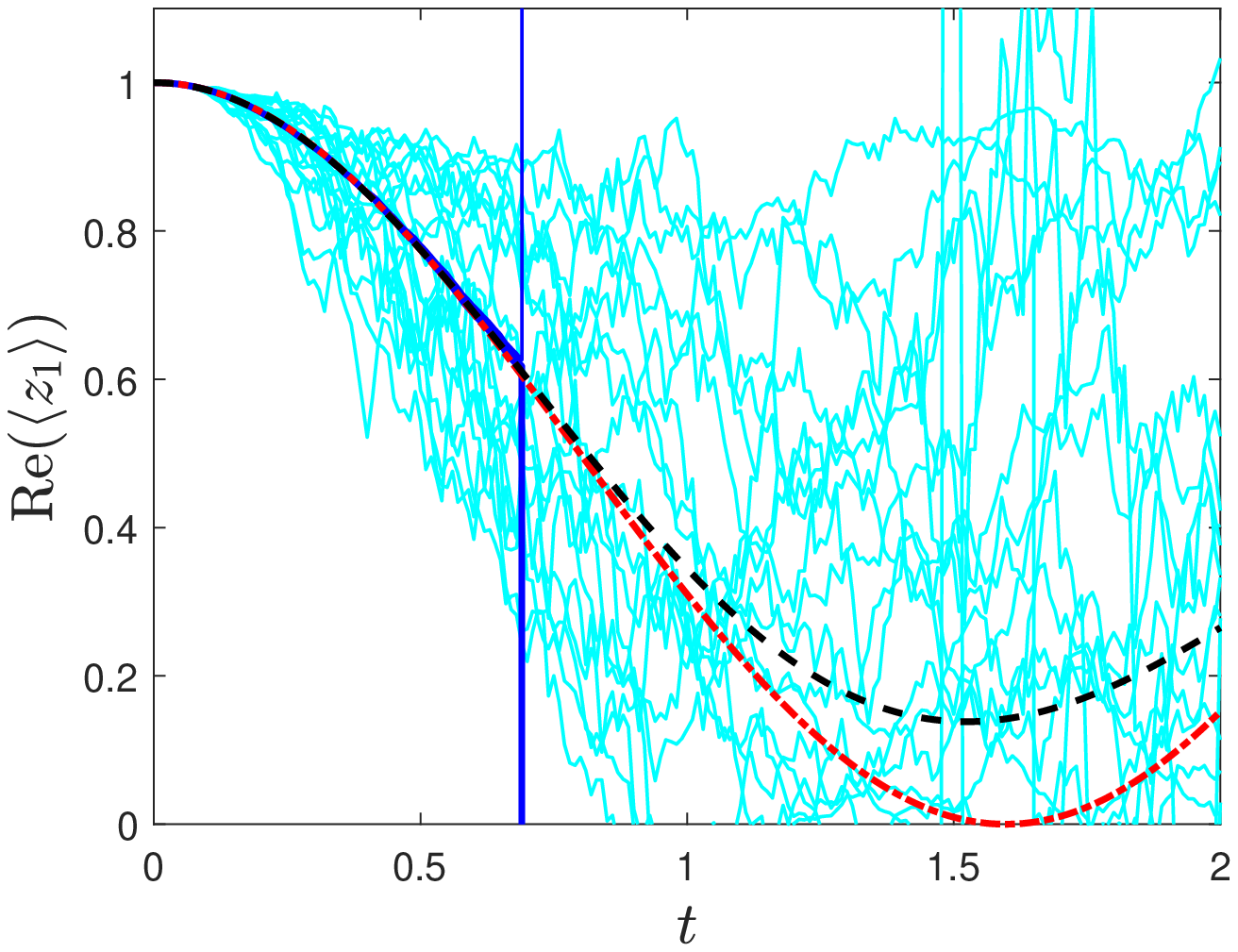}~\includegraphics[scale=0.62]{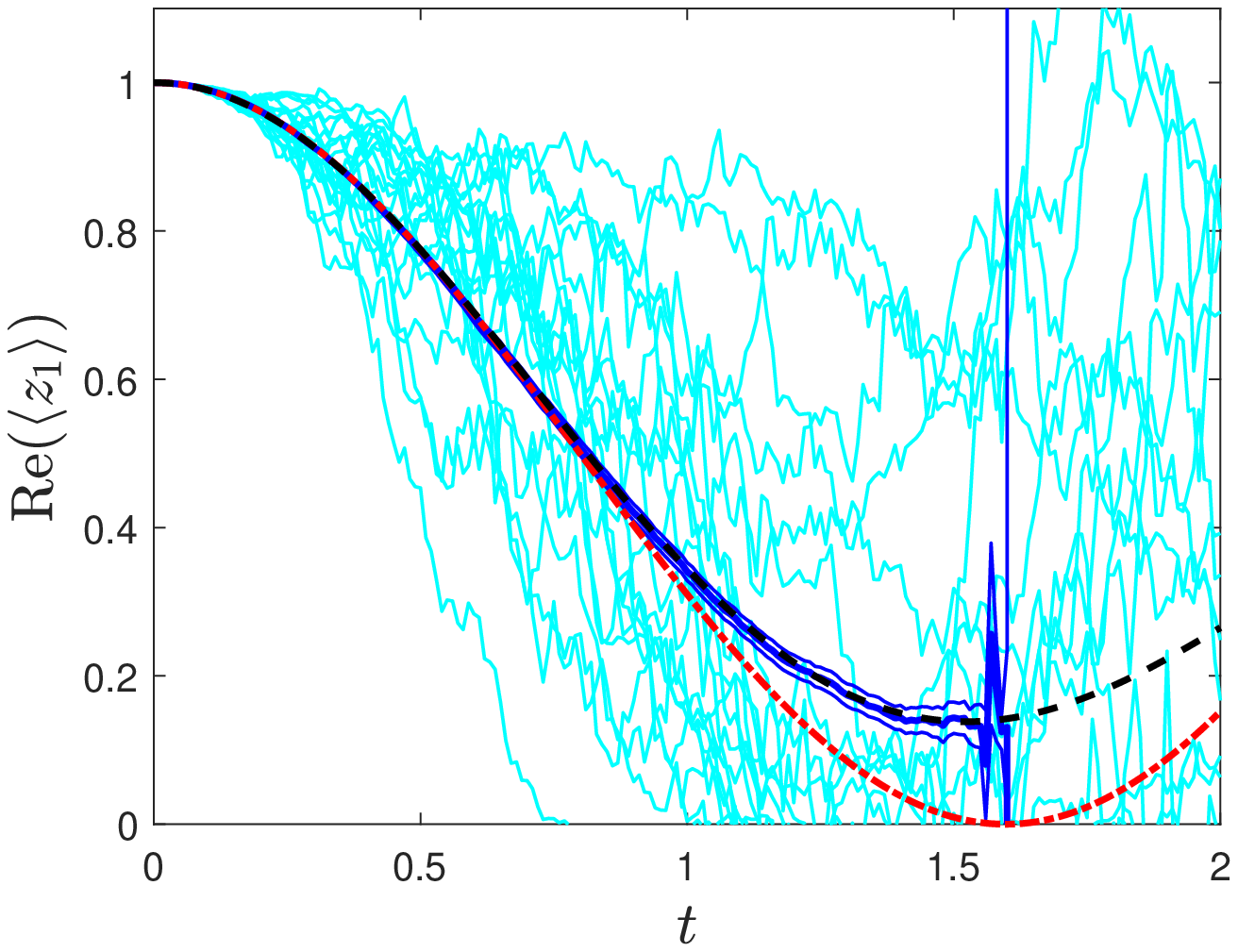}
\caption{(Colour online)
The occurrence of spikes in the stochastic dynamics for the Fermi--Hubbard model for individual trajectories with different diffusion gauges.
Left: Standard GPSR from the literature~\cite{Corboz_bookchapter_2013}, i.e., with the analytic noise matrix $B$ of equation~(\ref{eq:HubbAppBmatrix}).
Right:  GPSR with numerical noise matrices $B$, i.e., dynamically solving equation~(\ref{eq:D_BBT}).
Thin (cyan) curves are individual trajectories. Thick (blue) curves are the average of $10^3$ such trajectories, surrounded by $\pm$ the stochastic errors.
Dashed-dotted (red) curves are from the corresponding Hartree--Fock meanfield-method~\cite{RahavPRB2009}, while dashed (black) curves are from numerical exact diagonalization, for comparisons.
Parameters and initial conditions are $J=U=1$, respectively, according to~(\ref{eq:Hubbard_initial}).
}
\label{fig:spikes_for_hubbard}
\end{figure}
As initial condition $\vec{z}(0)$ for the Hamiltonian~(\ref{eq:HubbardH}), we choose
\begin{equation} \label{eq:Hubbard_initial}
\vec{z}(0)^T = \left[ 1, 0, 0, 0, 1, 0, 0, 0 \right],
\end{equation}
which corresponds to the initial ($t=0$) quantum-operator averages
\begin{equation} 
  \langle \hat{c}_{i,\sigma}^{\dagger}\hat{c}^{\phantom{\dagger}}_{j,\sigma}   \rangle =
  \begin{cases}
    1, \qquad &i = j = 1,\\
    0, &\text{otherwise}.
  \end{cases}
\end{equation}

We begin by examining the dynamics of the stochastic average $ \langle  z_1 \rangle $.
In figure~\ref{fig:spikes_for_hubbard}, we show $20$ individual trajectories and the stochastic averages over $10^{3}$ trajectories for the standard GPSR with analytic noise matrices (left panel) and the GPSR with numerical noise matrices (right panel).
The numerical diffusion gauges clearly outperform the analytic ones, yielding a practical simulation time that is twice as long.
Importantly, the average of the stochastic method follows the curve of the exact diagonalization, which differs from that of the Hartree--Fock meanfield-method.
On the contrary, we can hardly distinguish the three curves in the left panel of figure~\ref{fig:spikes_for_hubbard}.

\begin{figure}
\centering{}
\includegraphics[scale=0.7]{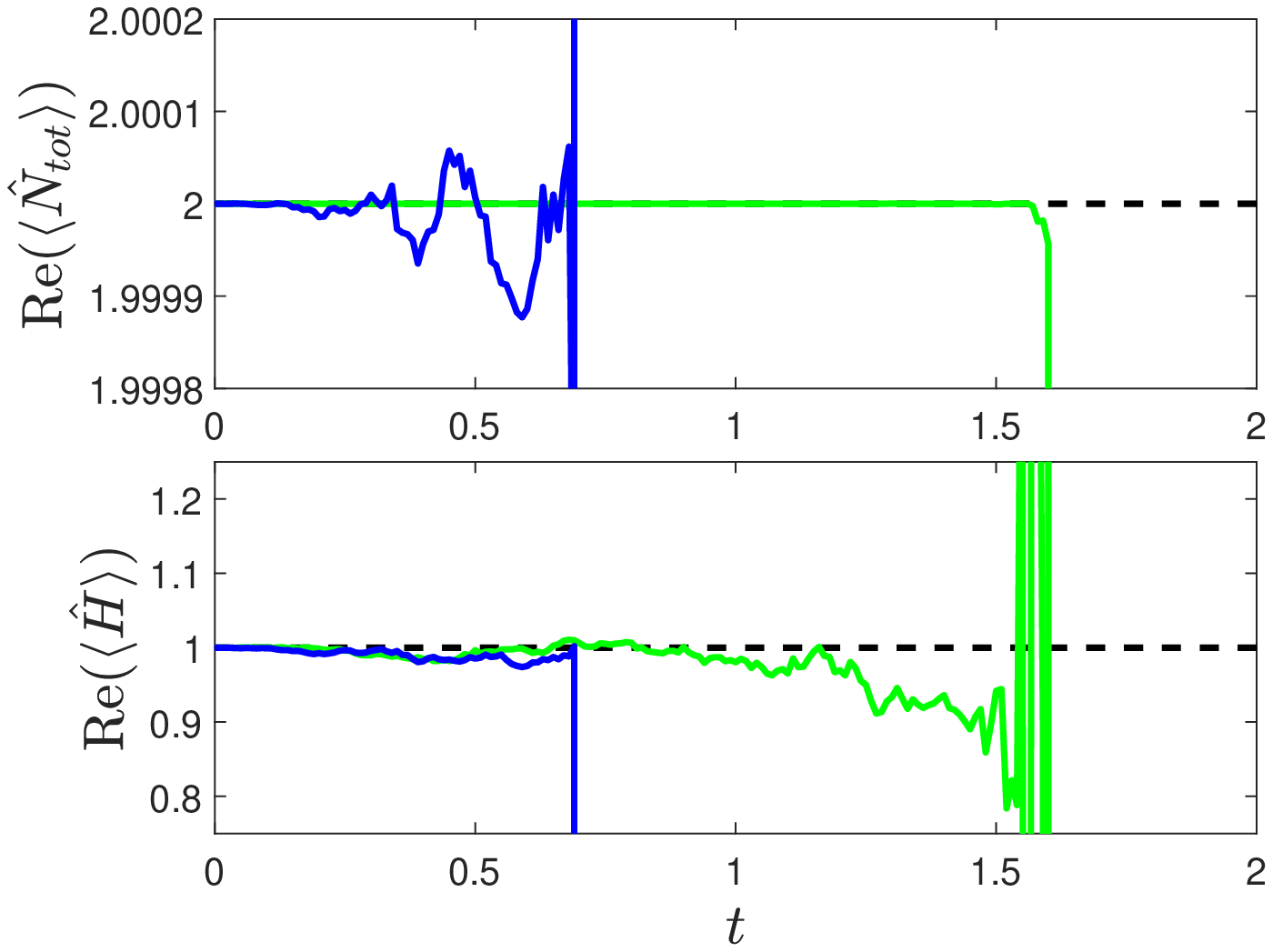}
\caption{(Colour online)
Benchmarking of time-dependent correlation functions for the Fermi--Hubbard model with different diffusion gauges.
Upper plot, total number of particles according to~(\ref{eq:Hubbard_N}).
Lower plot, total energy according to~(\ref{eq:Hubbard_H}).
Standard GPSR from the literature~\cite{Corboz_bookchapter_2013}, as in the left figure~\ref{fig:spikes_for_hubbard}, (blue) curves that here spike at $t \simeq 0.7$.
GPSR with numerical noise matrices $B$, as in the right figure~\ref{fig:spikes_for_hubbard}, (green) curves that here spike at $t \simeq 1.5$.
All four stochastic curves are averages of $10^3$ trajectories.
Dashed (black) curves are from exact diagonalization and are horisontal as expected.
Parameters and initial conditions are the same as in figure~\ref{fig:spikes_for_hubbard}.
}
\label{fig:observables_for_hubbard}
\end{figure}

We then investigate quantum correlations that map ($\leftrightarrow $) to stochastic averages over combinations of phase-space variables for the Fermi--Hubbard model.
In figure~\ref{fig:observables_for_hubbard}, we show in the top panel the total number of particles
\begin{equation} \label{eq:Hubbard_N}
 \langle \widehat{N}_{\textnormal{tot}} \rangle = \sum\nolimits _{j,\sigma}  \langle  \hat{n}_{{j},\sigma}  \rangle  \leftrightarrow   \langle z_1 \rangle + \langle z_4 \rangle + \langle z_5 \rangle + \langle z_8 \rangle ,
\end{equation}
and in the bottom panel the total energy
\begin{equation}
		 \langle \widehat{H} \rangle =  - J \sum\nolimits _{{\left\langle i,j\right\rangle },\sigma}  \langle \hat{c}_{{i},\sigma}^{\dagger}\hat{c}_{{j},\sigma} \rangle + U \sum\nolimits _{{j}}  \langle  \hat{n}_{{j},\uparrow}\hat{n}_{{j},\downarrow} \rangle
		 \leftrightarrow -J(\langle z_2 \rangle + \langle z_3 \rangle + \langle z_6 \rangle + \langle z_7 \rangle ) + U(\langle z_1 z_5 \rangle + \langle  z_4 z_8 \rangle).
\label{eq:Hubbard_H}
\end{equation}
We only report the real part of~(\ref{eq:Hubbard_N}) and~\eqref{eq:Hubbard_H}, as these two constants of motion are equal to the expectation values of the initial state~(\ref{eq:Hubbard_initial}), which are $\langle \widehat{N}_{\textnormal{tot}} \rangle=2$ and $\langle \widehat{H} \rangle=1$, respectively, and will thus have no imaginary part.

The Hartree--Fock method will produce a horizontal line (not shown) for the total energy, but this will be the sum of two terms that both deviate substantially from the correct values of kinetic- and interaction-energies.
%

\subsection{Results for the Fermi--Bose model}
%
\begin{figure}
\centering{}
\includegraphics[scale=0.62]{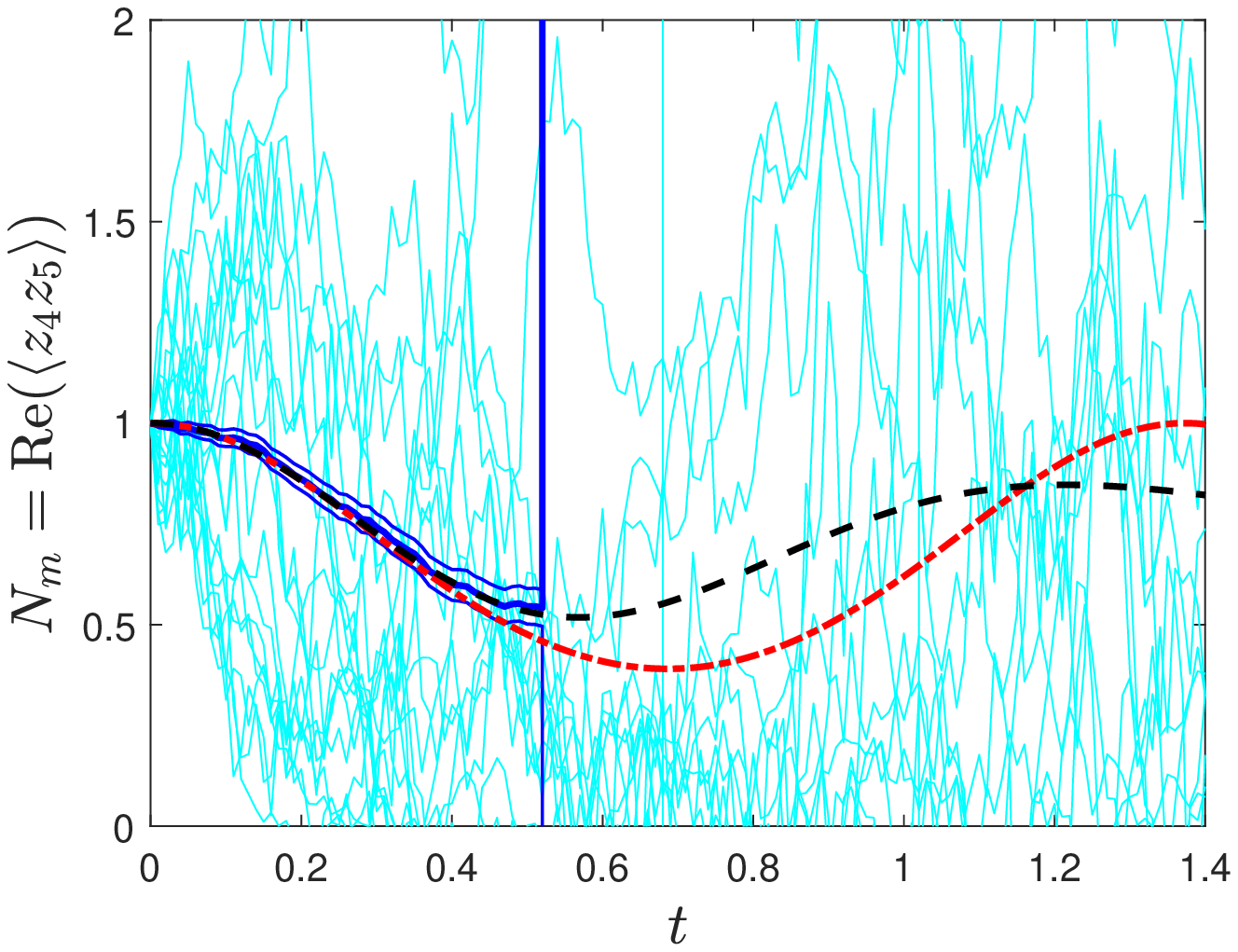}~\includegraphics[scale=0.62]{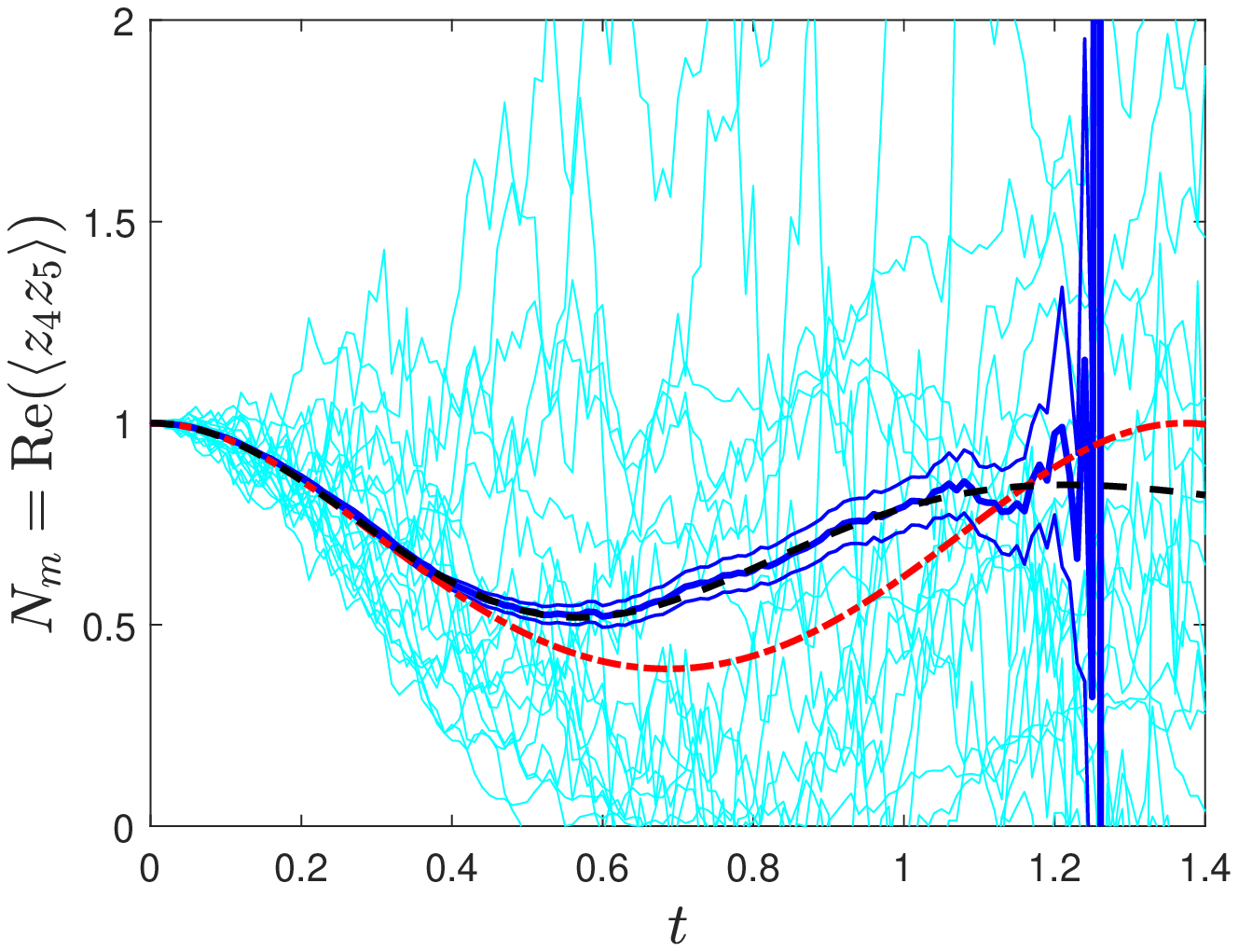}
\caption{(Colour online)
The occurrence of spikes in the stochastic dynamics for the Fermi--Bose model for individual trajectories with different diffusion gauges.
Left: Standard mixed phase-space representation from the literature~\cite{Ogren_EPL_2010}, i.e., with the analytic noise matrix $B$ of equation~(\ref{eq:ex_of_B_for_FBM}).
Right:  GPSR with numerical noise matrices $B$, i.e., dynamically solving equation~(\ref{eq:D_BBT}).
Thick (blue) curves is the average of $10^3$ such trajectories, surrounded by $\pm$ the stochastic errors.
Dashed-dotted (red) curves are from the corresponding pairing-meanfield-method~\cite{PMFT_Jack_and_Pu_PRA_2005}, while dashed (black) curves are from numerical exact diagonalization, for comparisons.
Parameters and initial conditions are $N_m(0)=\Delta_1=1$, and $\kappa=2$, respectively, according to~(\ref{eq:molecular_dissociation_initial_condition}).
}
\label{fig:spikes_for_FBM}
\end{figure}
As initial condition $\vec{z}(0)$ for the Hamiltonian~(\ref{eq:MBEC_H}), we choose
\begin{equation} \label{eq:molecular_dissociation_initial_condition}
\vec{z}(0)^T = \left[ 0, 0, 0, 1, 1 \right],
\end{equation}
i.e., $\sqrt{N_m(0)}=1$, which corresponds to the initial ($t=0$) quantum-operator averages
\begin{equation} 
\langle \hat{c}_{{1},\sigma}^{\dagger}\hat{c}^{\phantom{\dagger}}_{{1},\sigma}   \rangle =    \langle \hat{c}^{\phantom{\dagger}}_{{1},\uparrow} \hat{c}^{\phantom{\dagger}}_{{-1},\downarrow}   \rangle = 0 ,  \ \textnormal{and} \ \langle \hat{a}^{\dagger}\hat{a} \rangle = N_m(0),
\end{equation}
where the latter is the initial ($t=0$) number of dimer-molecules.

In figure~\ref{fig:spikes_for_FBM}, we show  $20$ individual trajectories and the stochastic average $N_m = \langle \hat{a}^{\dagger}\hat{a} \rangle \leftrightarrow \langle z_4 z_5 \rangle$ for the quantum dynamics of the number of molecules.
In this case, we compare the GPSR with numerical noise matrices (right panel) with the standard mixed-phase space representation from the literature (left panel), which uses an analytic noise matrix.
Again, we observe a doubling of the practical simulation time, and we remark that the average of the GPSR method in the right panel of figure~\ref{fig:spikes_for_FBM} follows the curve of the exact diagonalization, which differs from that obtained by the pairing-meanfield method.
Once again, the three curves are hardly distinguishable within the useful simulation time of the standard method, as is evidenced in the left panel of figure~\ref{fig:spikes_for_FBM}.

Next, we investigate the total number of particles
\begin{equation}
	\langle \widehat{N}_{\textnormal{tot}} \rangle = \sum_{k}  \langle  \hat{n}_{k}  \rangle + \langle \hat{a}_0^{\dagger} \hat{a}_0 \rangle  \leftrightarrow   \langle z_1 \rangle + \langle z_4 z_5 \rangle ,
	\label{eq:Fermi_Bose_N}
\end{equation}
and the total energy
\begin{equation}
		\langle \widehat{H} \rangle =   2 \sum_{k} \Delta_k \langle \hat{n}_{k} \rangle + i \kappa \sum_k \langle  \hat{a}_{0} \hat{m}_{k}^{\dagger} -  \hat{a}_{0}^{\dagger} \hat{m}_{k} \rangle
		\leftrightarrow 2 \Delta_1 \langle z_1\rangle + i \kappa ( \langle z_4 z_3 \rangle -  \langle z_5 z_2 \rangle ).
	\label{eq:Fermi_Bose_H}
\end{equation}

\begin{figure}
\centering{}
\includegraphics[scale=0.7]{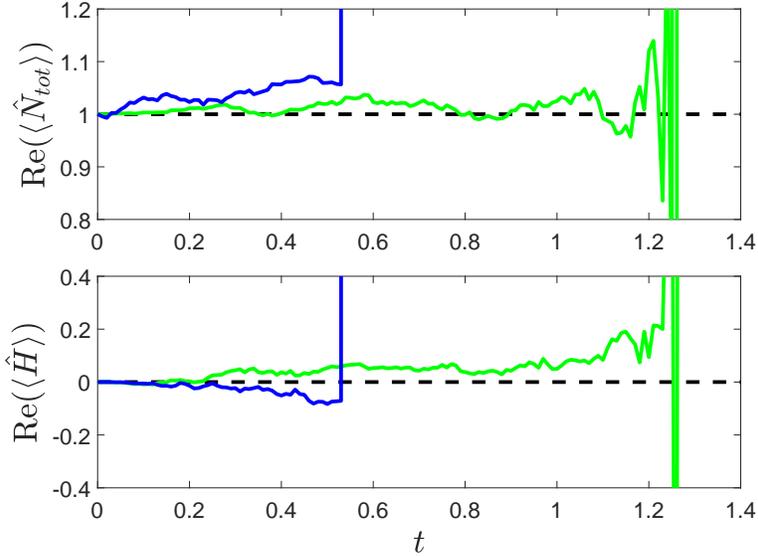}
\caption{(Colour online)
Benchmarking of time-dependent correlation functions for the Fermi--Bose model with different diffusion gauges.
Upper plot, total number of particles according to~(\ref{eq:Fermi_Bose_N}).
Lower plot, total energy according to~(\ref{eq:Fermi_Bose_H}).
Standard mixed phase-space representation from the literature~\cite{Ogren_EPL_2010}, as in the left figure~\ref{fig:spikes_for_FBM}, (blue) curves that here spike at $t \simeq 0.5$.
GPSR with numerical noise matrices $B$, as in the right figure~\ref{fig:spikes_for_FBM}, (green) curves that here spike at $t \simeq 1.2$.
Averages were calculated from $10^3$ trajectories,
parameters, and initial conditions are the same as in figure~\ref{fig:spikes_for_FBM}.
}
\label{fig:observables_for_FBM}
\end{figure}

The constants of motion are equal to the expectation values of the initial state~(\ref{eq:molecular_dissociation_initial_condition}), which are $\langle \widehat{N}_{\textnormal{tot}} \rangle=1$ (the atom with negative momentum is not included explicitly in the model) and $\langle \widehat{H} \rangle=0$ here.
We report the observed values of the real parts of these two quantities in figure~\ref{fig:observables_for_FBM}, as they are both expected to have no imaginary part.

For the total energy, the pairing-meanfield method will result in a horizontal line  (not plotted), but this will be the sum of two terms that both deviate substantially from the correct values of the kinetic- and conversion-energy.

\section{Summary and outlook}

Phase-space representations for quantum dynamics are an established tool for numerical simulations, but, often hampered by limited simulation times, they can only describe the initial dynamics of some large bosonic~\cite{Wuster_PRE_2017} and fermionic~\cite{Ogren_EPL_2010} quantum systems.
In order to unravel the exact quantum dynamics for larger times, it is essential to gradually increase the practical simulation time of these numerical computations.

Inspired by early work on analytic diffusion gauges for simple bosonic systems, we introduced a novel method based on the numerical evaluation of diffusion gauges, and we benchmarked the new technique against existing alternatives using two different complex fermionic systems.
For both the Fermi--Hubbard model~(figure~\ref{fig:spikes_for_hubbard}) and the Fermi--Bose model~(figure~\ref{fig:spikes_for_FBM}), we observed a doubling of the practical simulation time.

These numerical results are very encouraging, and they could, for example, trigger an investigation of the exact quantum dynamics of large 2D Fermi--Hubbard models.
Such models are currently beyond the reach of exact diagonalization, and have proven hard for matrix-product-states-based methods.

As for the reasons behind the prolonged simulation times observed, we can only speculate here.
It has been noted in the literature~\cite{Gilchrist_PRA_1997, Corboz_PRB_2008} that the emergence of spiking trajectories in the phase-space variables, i.e. a $|z_j|$ approaching infinite, can be connected to the build-up of fat tails in the distribution for $P\left(\vec{z}\right)$, and hence possible occurrence of boundary corrections.
Since the phase-space basis is overcomplete, alternative distributions $P\left(\vec{z}\right)$, and alternative dynamics of individual trajectories, may still describe the correct state.
With the introduced numerical diffusion gauges, which are different at each time step, the stochastic dynamics related to the distribution $P\left(\vec{z}\right)$ can be different, and in particular less likely to grow in the same directions in consecutive time steps.

In order to further increase the practical simulation time, research on a more fundamental level may be required.
The better we understand how these numerical diffusion gauges work, the more precise the constrained matrix equations we can construct to obtain them will be.

\section{Acknowledgements} 
We thank Joel Corney for useful discussions.

\appendix

\section{Motivations of the SDEs for the phase-space representations} \label{app:Derivation_of_SDEs}

In this appendix, we present one explicit instance of a system of SDEs for each of the three Hamiltonians in the main body of the article.
In each of the three examples, we keep the number of variables to a minimum.

\subsection{The positive-$P$ representation}
Now we explicitly present one possible system of SDEs for
the Hamiltonian~(\ref{eq:Bose-HubbardH}) with one bosonic mode.
Using the operator mapping for the bosonic positive-$P$ representation~\cite{GardinerBook1}, we obtain the $a_j$ and $D_{j,k}$ in equation~(\ref{eq:FPE}).
With $n=2$ phase-space variables ordered as $z_1=\alpha$, $z_2=\alpha^{+}$, the terms in the FPE~(\ref{eq:FPE}) with the first-order derivatives are~\cite{Steel_PRA_1998}
\begin{equation}
\left[\begin{array}{c}
a_{1}\left(\vec{z}\right)\\
a_{2}\left(\vec{z}\right)\end{array}\right]
=  -i
\left[\begin{array}{c}
\omega z_1 + \chi z_2 z_1^2\\
-\omega z_2 - \chi z_1 z_2^2\\
\end{array}\right].
\label{eq:FirstOrderDerivatives_positive_P}
\end{equation}
The second-order derivatives in the FPE~(\ref{eq:FPE}) are defined
by the symmetric $D \in \mathbb{C}^{2 \times 2}$ matrix
\begin{equation}
D = i \chi
\left[\begin{array}{cc}
d_{11} & 0\\
0 & d_{22}
\end{array}\right],
\label{eq:D_for_positive_P}
\end{equation}
where the two nonzero elements are
\begin{equation}
d_{11}=-z_1^2,\quad d_{22}= z_2^2.
\end{equation}
The matrix $B \in \mathbb{C}^{2 \times 2}$
\begin{equation}
B=
\sqrt{i \chi}
\left[\begin{array}{cc}
i z_1 & 0 \\
0 & -z_2
\end{array}\right],
\label{eq:ex_of_B_for_positive_P}
\end{equation}
then satisfies~(\ref{eq:D_BBT}) for $D$ in~(\ref{eq:D_for_positive_P}).
Hence, equation~(\ref{eq:FirstOrderDerivatives_positive_P}) together with equation~(\ref{eq:ex_of_B_for_positive_P}) give the SDE~(\ref{eq:SDE_Bose-Hubbard}).
However, the following one-parameter analytic $B$-matrix also satisfies (\ref{eq:D_BBT}) 
\begin{equation}
B(\gamma)=
-i \sqrt{i \chi}
\left[\begin{array}{rr}
\cosh(\gamma) z_1 & i \sinh(\gamma) z_1 \\
-\sinh(\gamma) z_2 & -i \cosh(\gamma) z_2
\end{array}\right].
\label{eq:ex_2_of_B_for_positive_P}
\end{equation}
The $B(\gamma)$ in~(\ref{eq:ex_2_of_B_for_positive_P}) is equivalent to the trigonometric form presented in~\cite{Deuar_PRA_2002} and was suggested for the positive-$P$ method for the Kerr oscillator in~\cite{Plimak_PRA_2001}.
For suitable values of the parameter, the noise matrix in~(\ref{eq:ex_2_of_B_for_positive_P}) shows a remarkable improvement in obtaining observables, as shown in figure~\ref{fig:observables_for_positive_P}, with lower variance for longer times, as shown in figure~\ref{fig:ex_1_for_positive_P}.
This improvement in the mean is due to the fact that the amplitudes $|z_1|$ and $|z_2|$ for the different trajectories, as seen in figures~2 and~3 in~\cite{Plimak_PRA_2001}.
For $\gamma = 0 $, equations (\ref{eq:ex_of_B_for_positive_P}) and~(\ref{eq:ex_2_of_B_for_positive_P}) are equal, while for $\gamma \simeq  3.1985$, corresponding to $\cosh(\gamma)=\sqrt{301/2}$ (see Appendix 2 of~\cite{Deuar_PRA_2002}) we get the improved result reported in figures~\ref{fig:ex_1_for_positive_P} and~\ref{fig:observables_for_positive_P} in section~\ref{sec:Introduction}.

\subsection{The two-site Fermi--Hubbard model}
In this subsection we explicitly present one possible system of SDEs for
the Hamiltonian (\ref{eq:HubbardH}) with two spatial sites.
Using the operator
mapping of the Gaussian phase-space representation \cite{Corney-fermionicII}
with only normal operator moments (no anomalous operator moments, i.e., no $m$ or $m^{+}$ variables are used here), one can derive a Fokker--Planck equation for the time-dependent probability distribution $P\left(\vec{z}\right)$ of the form~(\ref{eq:FPE}).

Here we use the notation $z_{1}=n_{11,\uparrow}$, $z_{2}=n_{12,\uparrow}$, $z_{3}=n_{21,\uparrow}$, $z_{4}=n_{22,\uparrow}$, $z_{5}=n_{11,\downarrow}$, $z_{6}=n_{12,\downarrow}$, $z_{7}=n_{21,\downarrow}$, $z_{8}=n_{22,\downarrow}$ for the eight complex phase-space variables we use for the two-site Fermi--Hubbard model.
The phase-space variables are related to the first-order physical moments $\left\langle \hat{n}_{ij,\sigma}\right\rangle \leftrightarrow  \langle n_{ij,\sigma} \rangle$, and their stochastic averages approach the quantum mechanical expectation values of those in the limit of many trajectories.

The terms in the FPE (\ref{eq:FPE}) with the first-order derivatives are in this case~\cite{Corboz_bookchapter_2013}
\begin{equation}
\left[\begin{array}{c}
a_{1}\left(\vec{z}\right)\\
a_{2}\left(\vec{z}\right)\\
a_{3}\left(\vec{z}\right)\\
a_{4}\left(\vec{z}\right)\\
a_{5}\left(\vec{z}\right)\\
a_{6}\left(\vec{z}\right)\\
a_{7}\left(\vec{z}\right)\\
a_{8}\left(\vec{z}\right)\end{array}\right] = i \left[\begin{array}{c}
J \left( z_2 - z_3 \right)\\
J \left( z_1 - z_4 \right) + U \left( z_5 - z_8 \right) z_2\\
J \left( z_4 - z_1 \right) + U \left( z_8 - z_5 \right) z_3\\
J \left( z_3 - z_2 \right)\\
J \left( z_6 - z_7 \right)\\
J \left( z_5 - z_8 \right) + U \left( z_1 - z_4 \right) z_6\\
J \left( z_8 - z_5 \right) + U \left( z_4 - z_1 \right) z_7\\
J \left( z_7 -z_6 \right) \end{array}\right].
\label{eq:HubbAppFirstOrderDerivatives}
\end{equation}
The second-order differential operator in~(\ref{eq:FPE}) are defined
by the symmetric $D \in \mathbb{C}^{8 \times 8}$ matrix has the form
%
\begin{equation} \label{app:eq_D_for_Hubbard}
D=\frac{iU}{2}\left[\begin{array}{cccccccc}
0 & 0 & 0 & 0 & 0 & d_{16} & d_{17} & 0\\
0 & 0 & 0 & 0 & d_{25} & d_{26} & d_{27} & d_{28}\\
0 & 0 & 0 & 0 & d_{35} & d_{36} & d_{37} & d_{38}\\
0 & 0 & 0 & 0 & 0 & d_{46} & d_{47} & 0\\
0 & d_{25} & d_{35} & 0 & 0 & 0 & 0 & 0\\
d_{16} & d_{26} & d_{36} & d_{46} & 0 & 0 & 0 & 0\\
d_{17} & d_{27} & d_{37} & d_{47} & 0 & 0 & 0 & 0\\
0 & d_{28} & d_{38} & 0 & 0 & 0 & 0 & 0\end{array}\right],
\end{equation}
where the twelve different types of nonzero elements are
\begin{equation}
  \begin{cases}
    d_{16}=2 z_6 \left(z_2 z_3 + z_1 \tilde{z}_{1} \right),\\
    d_{17}=-2 z_7 \left(z_2 z_3 + z_1 \tilde{z}_{1} \right),\\
    d_{25}=2 z_2 \left(z_6 z_7 + z_5 \tilde{z}_{5} \right),\\
    d_{26}=2 z_2 z_6 \left(z_4 + z_8 - z_1 - z_5 \right),\\
    d_{27}=2 z_2 z_7 \left(z_1 + z_8 - z_4 - z_5 \right),\\
    d_{28}=-2 z_2 \left(z_6 z_7 + z_8 \tilde{z}_{8}\right),\\
    d_{35}=-2 z_3 \left(z_6 z_7 + z_5 \tilde{z}_{5}\right),\\
    d_{36}=2 z_3 z_6 \left(z_4 + z_5 - z_1 - z_8 \right),\\
    d_{37}=2 z_3 z_7 \left(z_1 + z_5 - z_4 - z_8 \right),\\
    d_{38}=2 z_3 \left(z_6 z_7 + z_8 \tilde{z}_{8} \right),\\
    d_{46}=-2 z_6 \left(z_2 z_3 + z_4 \tilde{z}_{4} \right),\\
    d_{47}=2 z_7 \left(z_2 z_3 + z_4 \tilde{z}_{4} \right).
  \end{cases}
\end{equation}

One example that satisfies~(\ref{eq:D_BBT}) for~(\ref{app:eq_D_for_Hubbard}) is~\cite{Corboz_bookchapter_2013}
\begin{equation}
B=\sqrt{\frac{iU}{2}}\left[\begin{array}{cccccccc}
i\tilde{z}_{1} z_1 & -\tilde{z}_{1} z_1 & -i z_2 z_3 & z_2 z_3 & \tilde{z}_{1} z_1 & i \tilde{z}_{1} z_1 & - z_2 z_3 & -i z_2 z_3\\
-i z_1 z_2 & z_1 z_2 & i \tilde{z}_{4} z_2 & -\tilde{z}_{4} z_2 & \tilde{z}_{1} z_2 & i \tilde{z}_{1} z_2 & - z_2 z_4 & -i z_2 z_4\\
i\tilde{z}_{1} z_3 & -\tilde{z}_{1} z_3 & -i z_3 z_4 & z_3 z_4 & - z_1 z_3 & -i z_1 z_3 & \tilde{z}_{4} z_3 & i\tilde{z}_{4} z_3\\
-i z_2 z_3 & z_2 z_3 & i \tilde{z}_{4} z_4 & -\tilde{z}_{4} z_4 & -z_2 z_3 & -i z_2 z_3 & \tilde{z}_{4} z_4 & i\tilde{z}_{4} z_4\\
i\tilde{z}_{5} z_5 & \tilde{z}_{5} z_5 & -i z_6 z_7 & -z_6 z_7 & \tilde{z}_{5} z_5 & -i \tilde{z}_{5} z_5 & -z_6 z_7 & i z_6 z_7\\
-i z_5 z_6 & -z_5 z_6 & i \tilde{z}_{8} z_6 & \tilde{z}_{8} z_6 & \tilde{z}_{5} z_6 & -i \tilde{z}_{5} z_6 & -z_6 z_8 & i z_6 z_8\\
i \tilde{z}_{5} z_7 & \tilde{z}_{5} z_7 & -i z_7 z_8 & -z_7 z_8 & -z_5 z_7 & i z_5 z_7 & \tilde{z}_{8} z_7 & -i \tilde{z}_{8} z_7\\
-i z_6 z_7 & -z_6 z_7 & i \tilde{z}_{8} z_8 & \tilde{z}_{8} z_8 & -z_6 z_7 & i z_6 z_7 & \tilde{z}_{8} z_8 & -i \tilde{z}_{8} z_8
\end{array}\right].
\label{eq:HubbAppBmatrix}
\end{equation}
We remind the reader of the notation $\tilde{z}_{j}\equiv1-z_{j}$ for so-called hole-variables.
The $B \in \mathbb{C}^{8 \times 8}$ matrix (\ref{eq:HubbAppBmatrix}) is only one out of many possible ways to represent the second-order derivatives from the FPE (\ref{eq:FPE}) into an SDE~(\ref{eq:SDEs}).

\subsection{The mixed Fermi--Bose system for molecular dissociation}
In this subsection, we explicitly present one possible system of SDEs for
the Hamiltonian~(\ref{eq:MBEC_H}) with two fermionic momentum modes.
Using the operator mapping for the fermionic GPSR~\cite{Corney-fermionicII} combined with the bosonic positive-$P$ representation~\cite{GardinerBook1}, we obtain the $a_j$ and $D_{j,k}$ in equation~(\ref{eq:FPE}).
If we order the $n=5$ phase-space variables as $z_1=n_1$, $z_2=m_1$, $z_3=m_1^+$, $z_4=\alpha$, $z_5=\alpha^{+}$, then~(\ref{eq:FPE}) reduces to~\cite{Ogren_EPL_2010}
\begin{equation}
	\left[\begin{array}{c}
		a_{1}\left(\vec{z}\right)\\
		a_{2}\left(\vec{z}\right)\\
		a_{3}\left(\vec{z}\right)\\
		a_{4}\left(\vec{z}\right)\\
		a_{5}\left(\vec{z}\right)\end{array}\right]
	=
	\left[\begin{array}{c}
		\kappa\left( z_3 z_4 + z_2 z_5 \right)\\
		  -2i\Delta_1 z_2 + \kappa z_4 \left( 1 - 2 z_1 \right) \\
	  2i\Delta_1 z_3 + \kappa z_5 \left( 1 - 2 z_1 \right) \\
		-\kappa z_2\\
		-\kappa z_3\\
	\end{array}\right].
		\label{eq:FirstOrderDerivatives_molecular_dissociation} 
	\end{equation}

The second-order differential operator in the FPE~(\ref{eq:FPE}) is defined
by the symmetric $D \in \mathbb{C}^{5 \times 5}$ matrix
\begin{equation}
	D =
	\kappa
	\left[\begin{array}{ccccc}
		0 &  0 & 0 &  d_{14} & d_{15}\\
		0 &  0 & 0 &  d_{24} & d_{25}\\
		0 &  0 & 0 &  d_{34} & d_{35}\\
		d_{14} & d_{24} & d_{34} & 0 & 0\\
		d_{15} & d_{25} & d_{35} & 0 & 0
	\end{array}\right],
	\label{eq:D_for_FBM}
\end{equation}
where the six different types of nonzero elements are
\begin{equation}
  \begin{cases}
    d_{14}=z_1 z_2,\\
    d_{15}=z_1 z_3,\\
    d_{24}=z_2^2,\\
    d_{25}=-z_1^2,\\
    d_{34}=-z_1^2,\\
    d_{35}=z_3^2.
  \end{cases}
\label{eq:elements_of_D_for_molecular_dissociation}
\end{equation}

With the following $B \in \mathbb{C}^{5 \times 4}$ matrix
\begin{equation}
	B= 
	\frac{\sqrt{\kappa}}{\sqrt{2}}
	\left[\begin{array}{cccc}
		z_1 z_2 & -i z_1 z_2 & z_1 z_3 & -i z_1 z_3\\
		z_2^2 & -i z_2^2 & -z_1^2 & i z_1^2\\
		-z_1^2 & i z_1^2 & z_3^2 & -i z_3^2 \\
		1 & i & 0 & 0\\
		0 & 0 & 1 & i
	\end{array}\right],
	\label{eq:ex_of_B_for_FBM}
\end{equation}
we have one possible $n \times m$ matrix that solve equation~(\ref{eq:D_BBT}).
For a more compact explicit notation of the SDEs~(\ref{eq:SDEs}), we can first define a matrix $\widetilde{B}$ with half as many columns, $\widetilde{B}_{j,k} = B_{j,2k-1}$.

Then each of the $m/2$ columns of $\widetilde{B}$ is multiplied with by one of the $m/2$ complex noises $\xi_1, \ldots, \xi_{m/2}$, or by its complex conjugate if $\textnormal{Im}[B_{j,2k}] / \textnormal{Re}[B_{j,2k-1}] = -1$.
With a noise vector written with only 2 complex noises $\xi_1$ and $\xi_2$, according to equation~(\ref{eq:ComplexNoises}),
we have the noise terms from~\cite{Ogren_EPL_2010}, that is,
\begin{equation} \label{B_FBC}
	\vec{b} (\vec{z}, \vec{\xi}) = \sqrt{\kappa}
	\left[\begin{array}{c}
		z_1 z_2 \xi_1^* + z_1 z_3 \xi_2^*\\
		z_2^2 \xi_1^* - z_1^2 \xi_2^*\\
		-z_1^2 \xi_1^* + z_3^2 \xi_2^*\\
		\xi_1\\
		\xi_2
	\end{array}\right].
\end{equation}

It is the same procedure, but using 4 complex noises, that have been used in going from equation~(\ref{eq:HubbAppBmatrix}) to the SDE~(\ref{eq:SDE}) for the Fermi--Hubbard model.

\newcommand{\diag}{\ensuremath{\mathrm{diag}}}

\section{Numerical computation of $B$-matrices using Takagi’s decomposition}
We begin by recalling Takagi’s decomposition and the algorithms that can be  used to compute it.
We then describe how this decomposition can be used to construct a $B$-matrix satisfying~(\ref{eq:D_BBT}).

\begin{theorem}{\textnormal{(}Chapter 4, Corollary 4.4.4(c) of \cite{HoJo13}\textnormal{)}}
\label{td}
Let $D$ be a complex symmetric $n \times n$ matrix. Then there exist a unitary matrix $U$ and a nonnegative diagonal matrix $\varSigma = \diag(\sigma_1, \dots , \sigma_n)$ such that $D = U \varSigma U^T$.
\end{theorem}
Takagi’s decomposition can be interpreted as a symmetric analog of a Singular Value Decomposition (SVD) of $D$. Note that $\sigma_1, \dots , \sigma_n$ are, in fact, the singular values of $D$~\cite{Thom79}.

Several algorithms to compute Takagi's decomposition have been presented in the literature~\cite{GuQi03,WaLC18,XuQi08,BuGr88}.
In this paper, we use the divide-and-conquer method developed by Xu and Qiao \cite{XuQi08}. As this algorithm requires the input matrix be tridiagonal, we use the Lanczos tridiagonalization with modified partial orthogonalization and restart to reduce a complex symmetric matrix to a tridiagonal form \cite{GuQi03}.

By Theorem \ref{td}, we have that $D = U \varSigma U^T = \bigl( U \sqrt{\varSigma} \bigr) \bigl( U \sqrt{\varSigma} \bigr)^T$ for some unitary $U$ and nonnegative diagonal~$\varSigma$. If we define $B :=  U \sqrt{\varSigma}$, where $\sqrt{\varSigma} = \diag(\sqrt{\sigma_1}, \dots , \sqrt{\sigma_n})$, then it is easy to see that we have obtained the desired decomposition $D= B B^T$.
Note that this decomposition is not unique: for any orthogonal matrix $Q$, for example, we have $D=B Q Q^TB^T = (BQ)(BQ)^T =:  B^{\phantom{T}}_1 B_1^T$.

\section{It\^{o} calculus, to avoid Stratonovich corrections}

It is common to convert the It\^{o} stochastic differential equations to Stratonovich form and to integrate the latter with semi-implicit methods~\cite{Drummond_Mortimer_1991}, as these sometimes have better convergence properties.
This in general means that nonzero Stratonovich corrections (SC) to the drift-vector $\vec{a} (\vec{z})$, of the form
\begin{equation} \label{Stratonovich_correction}
SC_i = - \frac{1}{2} \sum_{j,k} B_{j,k} \frac{ \partial B_{i,k} }{\partial z_j},
\end{equation}
need to be added~\cite{GardinerBook1}.
These have been tested for the specific analytic examples of $B$-matrices here, i.e., equations~(\ref{eq:ex_of_B_for_positive_P}), (\ref{eq:HubbAppBmatrix}), and~(\ref{eq:ex_of_B_for_FBM}).
However, in the case of numerical $B$-matrices we want to avoid the numerical differentiations needed in~(\ref{Stratonovich_correction}), thus our chosen strategy was to only use It\^{o} calculus when applying numerical $B$-matrices.

\end{document}